\documentclass
[aps,prl,amsmath,amsfonts,amssymb,twocolumn,superscriptaddress,showpacs]{revtex4}%
\usepackage{times}
\usepackage{amsfonts}
\usepackage{amsmath}
\usepackage{amssymb}
\usepackage{graphicx}%
\providecommand{\U}[1]{\protect\rule{.1in}{.1in}}

\begin{document}
\title{Why Entanglement of Formation is not generally monogamous}

\author{F. F. Fanchini}
 \email{fanchini@iceb.ufop.br}
\affiliation{Departamento de F\'isica, Universidade Federal de Ouro Preto, CEP 35400-000, Ouro Preto, MG,
Brazil}
\author{M. C. de Oliveira}
 \email{marcos@ifi.unicamp.br}
\affiliation{Instituto de F\'{\i}sica Gleb Wataghin, Universidade Estadual de Campinas, CEP 13083-859,
Campinas, SP, Brazil}
\affiliation{Institute for Quantum Information Science, University of Calgary, Alberta T2N 1N4, Canada}
\author{L. K. Castelano}
\affiliation{Departamento de F\'isica, Universidade Federal de S\~ao Carlos, CEP 13565-905, S\~ao Carlos, SP, Brazil}
\author{M. F. Cornelio}
\affiliation{Instituto de F\'{\i}sica Gleb Wataghin, Universidade Estadual de Campinas, CEP 13083-859,
Campinas, SP, Brazil}

\begin{abstract}
Differently from correlation of classical systems, entanglement of quantum systems cannot be distributed at will - if one system $A$ is maximally entangled with another system $B$, it cannot be entangled at all to a third system $C$. This concept, known as the monogamy of entanglement, manifests when the entanglement of $A$ with a pair $BC$, can be divided as contributions of entanglement between $A$ and $B$ and $A$ and $C$, plus a term ${\tau}_{ABC}$ involving genuine tripartite entanglement and so expected to be always positive. A very important measure in Quantum Information Theory, the Entanglement of Formation (EOF), fails to satisfy this last requirement.  Here we present the reasons for that and show a set of conditions that an arbitrary pure tripartite state must satisfy for EOF to become a monogamous measure, ie, for ${\tau}_{ABC}\ge0$. The relation derived is connected to the \textit{discrepancy} between quantum and classical correlations, being ${\tau}_{ABC}$ negative whenever the quantum correlation prevails over the classical one. This result is employed to elucidate features of the distribution of entanglement during a dynamical evolution. It also helps to relate all monogamous instances of EOF to the Squashed Entanglement, an always monogamous entanglement measure.
\end{abstract}
\date{\today}
 \maketitle

\section{Introduction}
Monogamy of an entanglement measure $\cal{E}$ asserts that, in a tripartite $A$, $B$ and $C$ system, the entanglement of $A$ with $BC$ can be divided as ${\cal{E}}_{A|BC}={\cal{E}}_{A|B}+{\cal{E}}_{A|C}+{\tau}_{ABC}$, where ${\cal{E}}_{A|i}, i=B,C$, is a bipartite entanglement and ${\cal{E}}_{ABC}$ is a genuine tripartite entanglement.  In that sense, differently from correlation in classical systems, there is a trade-off between the amount of bipartite entanglement $A$ can share with $B$ and $C$. About ten years ago Coffman, Kundu and Wootters (CKW) \cite{coffman} derived a monogamous relation for the squared concurrence and defined  the genuine tripartite entanglement as \textit{tangle} (hereafter called \textit{concurrence tangle}) \cite{coffman},
$\tau_{ABC} = C^2_{A(BC)} - C^2_{AB} - C^2_{AC}$, where $C^2_{ij}$ is the square of the concurrence between the pair $i$ and $j$. The concurrence tangle is always positive for a three-qubit system \cite{coffman} and for  multi-qubit systems \cite{tobias}. However, it is noticed that a similar analysis made with the EOF would give a tangle that can be positive or negative (hereafter we call this tangle as the \textit{EOF tangle}).   Although there are some instances in which EOF  could be distributed in a monogamous fashion, it is known that it is not, in general, a necessarily monogamous entanglement measure (See a more complete discussion in Refs. \cite{tro,marcio,barry1}). This is puzzling, since the EOF satisfies many of the axioms required for a good entanglement measure and, further, has a clear operational meaning \cite{benn}. So why is the EOF tangle negative or positive?

In fact it is now known that entanglement is not the only form of quantum correlations, since there are instances where a state which is separable (not entangled) still possesses a sort of correlation which can be used to perform certain tasks more efficiently than with classical correlation only. It is not surprising though that both forms of quantum correlations can be related to each other through extended system \cite{footnote1} distribution formulas. For example, it is possible to describe a conservation relation \cite{law} for distribution of EOF and Quantum Discord (QD), a measure of quantum correlation - for an arbitrary tripartite pure system, the sum of the QD of a chosen partition, given measurements on the complementary partitions, must be equal to the sum of the pairwise Entanglement of Formation (EOF) between the chosen partition and the complementary ones. Surprisingly, the sum of the pairwise EOF appears in a quite similar fashion to the desired expression for the so-called monogamy of entanglement, and the relation obtained can be connected to the way that classical correlations \cite{hen,conjec} are distributed \cite{LII}.

Until few years ago, the conjecture that the classical correlation would always be greater than the quantum correlation for any quantum state was broadly accepted \cite{hen,conjec}. In 2009, Maziero et. al \cite{schange} presented the first counterexample to this conjecture, while  studying the dissipative dynamics for two qubits. Despite their findings, the balance between classical and quantum correlation has never been connected to any quantum measure or protocol. 
In this paper, we show necessary and sufficient conditions for the monogamy of EOF to be established with the help of a general quantum correlation measure,  the QD, and identify the EOF tangle for an arbitrary tripartite state as the difference between classical and quantum correlations. To the best of our knowledge, it is the first time that the balance between classical and quantum correlation is shown to be crucial to understand this important open problem. For that we develop an operational interpretation to the EOF tangle as a measure of the quantum and the classical correlations unbalance, here called correlation {\textit{discrepancy}}.  Specific bounds are given in terms of conditional entropies, and the general results are discussed  through examples for states in  $2\times2\times 2$  and  $2\times2\times N$-dimensional Hilbert spaces. {Moreover}, the monogamic relation for the EOF helps to understand when this measure of entanglement is connected with the Squashed Entanglement \cite{chrwinter}. We give in a simple picture an interesting connection of monogamic instances of both EOF and QD with the Squashed Entanglement for arbitrarily tripartite mixed states. The paper is divided as follows. In Sec. II we derive the main relations for understanding the monogamic instances of the EOF. In Sec. III we extend the EOF tangle to a permutation invariant form and derive some important results for classes of states in $2\times2\times 2$  and for  $2\times2\times N$-dimensional Hilbert space in general. In Sec. IV we employ the results to understand the interplay of bipartite and multipartite entanglement in the sudden death of entanglement phenomenon. In Sec. V we sketch an extension for arbitrarily mixed states and show how monogamic instances of the EOF are connected to the Squashed Entanglement. Finally in Sec. VI a discussion encloses the paper.
\section{Conditions for Monogamy of EOF}
Although there is a common sense that correlation of classical systems can be distributed at will \cite{coffman}, a direct quantification of that is necessary. When thinking of correlation between two stochastic variables, $X$ and $Y$, the mutual entropy of Shannon $H(X:Y)\equiv H(X) + H(Y) - H(X,Y)$ is the optimal measure. It turns out that the mutual information is not always subadditive, \textit{i.e.}, $H(X:Y,Z)\not\le H(X:Y)+H(X:Z)$. But now let us suppose that we can increase $H(X:Y)$ to be maximal  (with $H(X)\le H(Y)$) so that  $H(X:Y)=H(X)$. It means that $H(X:Z)=H(X:Y)-H(X|Z)$, and so is increased linearly with $H(X:Y)$, being only constrained by $H(X:Z)\le H(X:Y)$. When extended to quantum system, in terms of the von Neumann entropy, the mutual information writes $S(A:B)\equiv I_{AB}=S_A + S_B - S_{AB}$ and the best we can do to infer on the distributon of correlation is to write $S(A:C)=S(A:B)+S(A|B)-S(A|C)$, with surprising consequences, since the conditional entropy can be negative whenever there is some entanglement in the parties involved. For, example if the joint system $ABC$ is pure, then $S(A:C)=S(A:B)+2S(A|B)=S(A:B)-2S(A|C)$.   Also $S(A:B,C)\not\le S(A:B)+S(A:C)$, being nonetheless additive if $ABC$ is pure. The mutual information quantifies both classical and quantum correlation. 

{In a different point of view}, the classical correlation \cite{hen,qd}  in a quantum state can be captured by
\begin{equation}
J_{AB}^\leftarrow =\max_{\left\{ \Pi
_{k}\right\} }\left[S_A-\sum_{k}p_{k}S_{A|k}\right],\label{classical}
\end{equation}
where $S_{A|k}$ is the conditional entropy after a measurement in $B$. Explicitly $S_{A|k}\equiv S(\rho_{A|k})$, where $\rho_{A|k}=\mathrm{Tr}_{B}(\Pi _{k}\rho_{AB} \Pi _{k})/\mathrm{Tr}_{AB}(\Pi_{k}\rho_{AB}\Pi_{k})$ is the reduced state of $A$
after the outcome $k$ in $B$ and $\{\Pi_k\}$ is a complete set of positive operator valued measurement resulting in the outcome $k$ with probability $p_k$. Since a measurement might give different results depending on the basis choice, a maximization over $\{\Pi_k\}$ is required. Thus $J_{AB}^\leftarrow$ is the maximal locally accessible mutual information, or the maximum amount of $AB$ mutual information that one can get by only performing local measurement on $B$ \cite{demon}. On the other hand, quantum discord, as a measure of quantum correlation, gives the amount of locally inaccessible mutual information, and so it is simply defined as
\begin{equation}
\delta_{AB}^\leftarrow = I_{AB} - J_{AB}^\leftarrow.\label{discordia}
\end{equation}
The equation above gives the amount of information that is not locally accessible by measurements on $B$ \cite{demon}. It is important to emphasize that the quantum discord is asymmetric, since the amount of locally inaccessible information for $B$ can be different from the locally inaccessible information for $A$. This means that generally $\delta^\leftarrow_{AB} \ne \delta^\leftarrow_{BA}$. Since $I_{AB}$ can be regarded as the mutual information previous to a measurement, while $J_{AB}^\leftarrow$ is the mutual information accessed by local measurements in $B$, we can identify $\delta^\leftarrow_{AB}$ as a measure on how much a state $\rho_{AB}$ is affected by local measurements on $B$.

It is also convenient to define a new quantity measuring the quantum and the classical correlations unbalance, or the \textit{correlation}    {\textit{discrepancy}},  or simply \textit{discrepancy} as\begin{equation}
\Delta^\leftarrow_{AB}\equiv J_{AB}^\leftarrow - \delta^\leftarrow_{AB}.\label{disharmony}
\end{equation}
This new measure has the advantage of having a clear meaning - It is the balance between the locally accessible mutual information and the locally inaccessible mutual information \cite{LII}. Similarly to quantum discord and classical correlation, the {\textit{discrepancy}}  is an asymmetric quantity and, contrarily to the formers, it can be negative or positive. Actually, if $\Delta^\leftarrow_{AB}>0$ the amount of classical correlation is larger than the quantum one, and vice versa. The {\textit{discrepancy}} is bounded by the mutual information as $-I_{AB}\le \Delta^\leftarrow_{AB}\le I_{AB}$. Moreover, {\textit{discrepancy}} is strictly zero if the bipartite quantum state is pure since, in that case, the quantum and classical correlations are both equal to $S_A$ (see the illustration in Fig. (\ref{dis})).
{It is interesting to interpret the discrepancy in terms of classical and quantum Maxwell demons \cite{demon, maxwell}.  Quantum demons have the ability to extract more work from correlated systems, in comparison to their classical counterpart, since they have access to nonlocal observables. The difference of their efficiencies is given by the quantum discord which, so, is always positive. The discrepancy compares this gain with the efficiency of classical demons.
If the gain given by the use of global operations do not exceed the work extracted by the local ones, the discrepancy is positive, and it is negative otherwise. In that case,  EOF is monogamous as we will see bellow. So, in this sense, the discrepancy can be seen as a gain of information when global operations are employed in contrast to local ones.
 Indeed by writing $\Delta^\leftarrow_{AB}=I_{AB}-2\left[S(A|B)-S_q(A|B)\right]$, where $S(A|B)$ is the conditional entropy  and $S_q(A|B)=\min_{\left\{ \Pi
_{k}\right\} }\sum_{k}p_{k}S_{A|k}$, clarifies the meaning of this interpretation.

We can yet compute the   {\textit{discrepancy}}  $\Delta^\leftarrow_{AB}$ as a function of the EOF and the mutual information. Using Eq. (\ref{classical}) it is straightforward to show that $\Delta^\leftarrow_{AB}=I_{AB} - 2\delta^\leftarrow_{AB}$. Moreover $\delta^\leftarrow_{AB}$ is related to the EOF of the pair $AC$, $E_{AC}$, through the relation \cite{law,koashi2004} $\delta^\leftarrow_{AB}=E_{AC}+S_{A|C}$. Those relations, when properly combined, give the   {\textit{discrepancy}}  as a function of the EOF, $\Delta^\leftarrow_{AB}=I_{AC}-2E_{AC}$. This equation shows that the   {\textit{discrepancy}}  between the pair $AB$ is positive if $I_{AC}/2>E_{AC}$ and relates the balance between the classical and the quantum correlations to the balance of the entanglement and the mutual information. In fact $I_{AC}/2$ is a lower bound to the entanglement of purification \cite{horodivi} and so, whenever the EOF is smaller than that, the information gain on using global operations does not exceed the information acquired with local operations only.  Finally, the  {\textit{discrepancy}}  also follows a conservation relation - Observing the   {\textit{discrepancy}}  as a function of the EOF and the mutual information, and noting that these quantities are symmetric, \textit{i.e.} $E_{AC}=E_{CA}$ and $I_{AC}=I_{CA}$, we obtain
\begin{equation}
\Delta^\leftarrow_{AB}=\Delta^\leftarrow_{CB}.\label{conserv}
\end{equation}
This equality means that in a pure tripartite  joint system \textit{the discrepancy of the correlations established with a particular system is constant under a permutation of the complementary subsystems}, or that the gain of information on using global operations over local operations is constant under permutation of the complementary subsystems. This fundamental aspect is general,  valid for any tripartite pure state, and rules how the classical and quantum correlations are distributed among the parties.
\begin{figure} 
\begin{center}
\includegraphics[width=.47\textwidth]{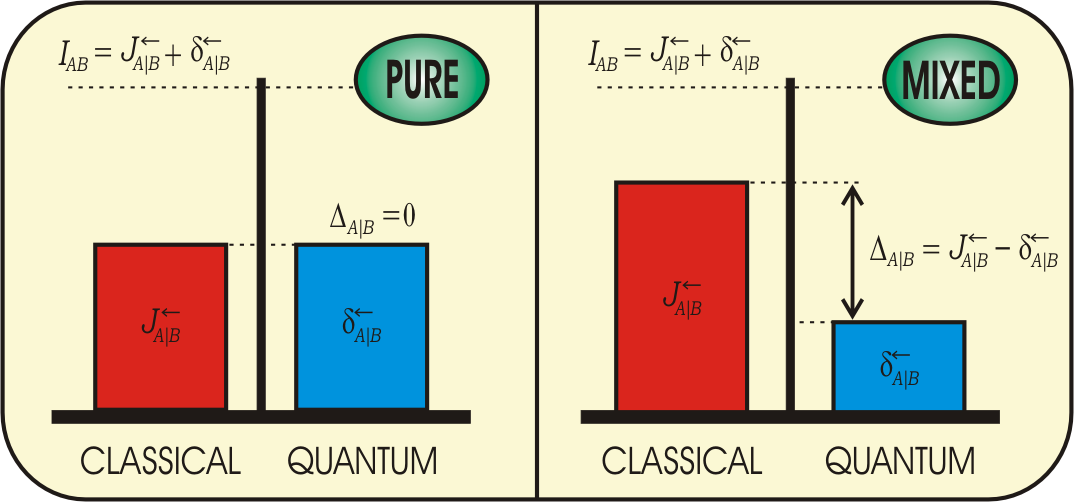} {}
\end{center}
\caption{(Color Online) An illustrative scheme of the   {\textit{discrepancy}}  for pure and mixed states. Here, for mixed state, we just considered in the illustration $J_{AB}^\leftarrow > \delta_{AB}^\leftarrow$, but the opposite is also possible.}\label{dis}
\end{figure}

Given the definitions above 
we now discuss on the condition for a monogamous distribution of EOF through the system. We begin by using the conservation law for distribution of EOF and QD \cite{law}
\begin{equation}
E_{AB}+E_{AC}=\delta^\leftarrow_{AB} + \delta^\leftarrow_{AC}, \label{law}
\end{equation}
being $E_{ij}$ the EOF between subsystems $i$ and $j$. Eq. (\ref{law}) can be rewritten in a different form by considering the EOF between $A$ and the joint system $BC$. Since the tripartite system is pure, we have that $E_{A(BC)}=S_A$, the von Neumann entropy of the subsystem $A$. Thus, we can write the relation for distribution of the EOF as
\begin{equation}
E_{AB}+E_{AC}+ \tau_A = E_{A(BC)}, \label{mono1}
\end{equation}
where the EOF tangle $\tau_A$ is identified as
\begin{equation}
\tau_A\equiv S_A - \delta^\leftarrow_{AB} - \delta^\leftarrow_{AC}.\label{tau1}
\end{equation}
Although the above equation already gives the required tangle of the EOF, it does not elucidate the operational meaning of $\tau_A$, which is apparent when we relate it to the classical correlation. Substituting Eq. (\ref{discordia}) into Eq. (\ref{tau1}), we find that
$\tau_A = J_{AB}^\leftarrow + J_{AC}^\leftarrow - S_A$, which combined with Eq. (\ref{tau1})
results in
\begin{equation}
\tau_A= \frac{1}{2}[\Delta^\leftarrow_{AB} + \Delta^\leftarrow_{AC}].\label{tau}
\end{equation}
This equation shows that the EOF tangle is the average of the correlation   {\textit{discrepancy}}, thereby giving the unbalance between the classical and quantum correlation. The EOF tangle written in such a form provides the desired operational interpretation in terms of the quantum and classical correlation balance. Moreover, based on this result, it is straightforward to predict when the EOF tangle (Eq. (\ref{tau})) is negative or not, \textit{i.e.} 
\begin{equation}
\tau_A\ge0\;\;\;\;\;\textrm{if and only if}\;\;\;\;\;J_{AB}^\leftarrow + J_{AC}^\leftarrow \ge \delta^\leftarrow_{AB} + \delta_{AC}^\leftarrow.\label{nove}
\end{equation}
So monogamy of EOF is equivalent to the inequality
$J_{AB}^\leftarrow + J_{AC}^\leftarrow \ge \delta^\leftarrow_{AB} + \delta_{AC}^\leftarrow$.
Explicitly Eq. (\ref{nove}) tells that whenever the minimized quantum conditional entropies $S_q(A|i)= \min_{\left\{ \Pi
_{k}\right\} }\sum_{k}p_{k}S(\rho_{A|k})$,  $\rho_{A|k}=\mathrm{Tr}_{i}(\Pi _{k}^i\rho_{Ai} \Pi _{k}^i)/\mathrm{Tr}_{Ai}(\Pi_{k}^i\rho_{Ai}\Pi_{k}^i)$, $i=B,C$ follow the inequality $S_A\ge S_q(A|B)+S_q(A|C)$, the EOF is monogamous. Note that each of the quantum conditional entropies is upper bounded by $S_A$ and so the above is a more restrictive bound. Thus whenever
\begin{equation}
S_A< S_q(A|B)+S_q(A|C)\le 2 S_A,
\end{equation}
EOF is \underline{not} a monogamous measure and $\tau_A$ is negative, meaning that the use of global operations gives a information gain which is larger than the information acquired with local operations only.
\section{permutation invariant tangle and applications}
One can note that the EOF tangle is not invariant under permutations of the subsystems, once $\tau_A\ne\tau_B\ne\tau_C$ in general, for $ \tau_B= (\Delta^\leftarrow_{BA} + \Delta^\leftarrow_{BC})/2$, and $ \tau_C= (\Delta^\leftarrow_{CB} + \Delta^\leftarrow_{CA})/2$, in contrast to the concurrence tangle \cite{coffman}. The reason for that is simple since the EOF tangle is not a measure a genuine tripartite entanglement, being a measure of the gain on using global operations over local ones instead. The different $ \tau_A$, $ \tau_B$ and $ \tau_C$ are related to difference instances where the local operations are made.
To define an invariant quantity and an essential measure of the global gain in the triple, we use the sum of the distinct EOF tangles.
To calculate the sum of the EOF tangles, we use the notation in terms of the flow of quantum correlation \cite{LII}, defined as
\begin{eqnarray}
{\cal{L}}_{\circlearrowright}&\equiv&\delta^{\leftarrow}_{BA}+\delta^{\leftarrow}_{CB}+\delta^{\leftarrow}_{AC},\label{clock}\\
{\cal{L}}_{\circlearrowleft}&\equiv&\delta^{\leftarrow}_{CA}+\delta^{\leftarrow}_{BC}+\delta^{\leftarrow}_{AB},\label{countclock}
\end{eqnarray}
and the flow of classical correlation
\begin{eqnarray}
{\cal{J}}_{\circlearrowright}&\equiv&J^{\leftarrow}_{BA}+J^{\leftarrow}_{CB}+J^{\leftarrow}_{AC},\label{Jclock}\\
{\cal{J}}_{\circlearrowleft}&\equiv&J^{\leftarrow}_{CA}+J^{\leftarrow}_{BC}+J^{\leftarrow}_{AB}.\label{Jcountclock}
\end{eqnarray}
In Eqs. (\ref{clock}) and (\ref{Jclock}) the correlations between pairs are computed sequentially as $A\rightarrow B\rightarrow C$, \textit{i.e.}, for the tripartite system $ABC$, we start by doing measurements on $A$ to infer the correlations (quantum and classical) between $A$ and $B$, then we measure $B$ to infer the correlations between $B$ and $C$, and finally we measure $C$ to infer the correlations between $C$ and $A$, representing a clockwise, ${\cal{L}}_{\circlearrowright}$ and ${\cal{J}}_{\circlearrowright}$, flow of information. Reversely, the computation of the correlations in Eqs. (\ref{countclock}) and (\ref{Jcountclock}) for the sequence $C\rightarrow B\rightarrow A$ represents a counterclockwise flow. We now recall a result given in our previous work\cite{LII}. For a general tripartite pure state, ${\cal{L}}_{\circlearrowright}={\cal{L}}_{\circlearrowleft}$, and thus through Eq. (\ref{classical}) together with the fact that the mutual information is symmetric, $I_{ij}=I_{ji}$, it is straightforward to prove that ${\cal{J}}_{\circlearrowright}={\cal{J}}_{\circlearrowleft}$. So we can write the sum of the EOF tangle as,
\begin{equation}
\tau_{ABC}\equiv\tau_A+\tau_B+\tau_C={\cal{J}}_{\circlearrowright}-{\cal{L}}_{\circlearrowright}=\Delta_{\circlearrowright}, \label{tauabc}
\end{equation}
where $\Delta_{\circlearrowright}\equiv \Delta^{\leftarrow}_{BA}+\Delta^{\leftarrow}_{CB}+\Delta^{\leftarrow}_{AC}$. This equation proves that the flow of the locally accessible information (classical correlation) minus the flow of the inaccessible information (quantum discord), or the clockwise flow of {\textit{discrepancy}}, is equal to the sum of the EOF tangle. 
Indeed $\tau_{ABC}$ is positive whenever the information gain through the usage of global operations is not larger than the information acquired with local operations on every party. It is important to remark that this feature is general and valid for arbitrary tripartite pure states. 

As an example, we calculate $\tau_{ABC}$ for many states belonging to the two 
 inequivalent \textit{families} of three-qubits entangled states, called GHZ and W, that can not be converted into each other by local operations and classical communication \cite{family}:
\begin{eqnarray}
&&|GHZ\rangle = \theta|\uparrow\uparrow\uparrow\rangle + \phi|\downarrow\downarrow\downarrow\rangle,\label{estghz}\\
&&|W\rangle= \alpha|\uparrow\uparrow\downarrow\rangle + \beta|\uparrow\downarrow\uparrow\rangle + \gamma|\downarrow\uparrow\uparrow\rangle.\label{estw}
\end{eqnarray}
 For the GHZ state defined by Eq. (\ref{estghz}), it is direct to show that $\tau_{ABC}>0$, because the QD between any bi-partition is always zero. For $\theta=\phi$, we have that ${\tau_{ABC}}=1$. 
  For the $W$ state, Eq.~(\ref{estw}), the situation is quite different because $J^\leftarrow_{ij}\ne\delta^\leftarrow_{ij}\ne0$ and ${\tau_{ABC}}=-0.546$, when $\alpha=\beta=\gamma$. In fact, we can show that $\tau_{ABC}<0$ for any $W$ state written in the form described in Eq.~(\ref{estw}). To prove this, we present an analytical computation of the EOF tangles $\tau_A$, $\tau_B$ and $\tau_C$ for an arbitrary three-qubit state.
We begin by using the definition of the {\textit{discrepancy}} as a function of the EOF and the mutual information, $\Delta^\leftarrow_{AB}=I_{AC}-2E_{AC}$. 
For the $W$ state, the EOF and the mutual information for any bipartition, $AB$, $BC$, and $CA$ can be easily computed and depend only on the modulus of $\alpha$, $\beta$, and $\gamma$, where $\alpha=\sin(\theta)\cos(\phi)$, $\beta=\sin(\theta)\sin(\phi)$, and $\gamma=\cos(\theta)$. In Fig. (\ref{tauW}), we show $\tau_{ABC}$ as a function of $\theta$ and $\phi$ and we observe that the EOF tangle is always negative for all $W$ state described by Eq.~(\ref{estw}). Therefore, we can understand the signal of $\tau$ as specifying which type of tripartite entanglement we have, W or GHZ.
\begin{figure}[htbp]
\begin{center}
\includegraphics[width=.5\textwidth]{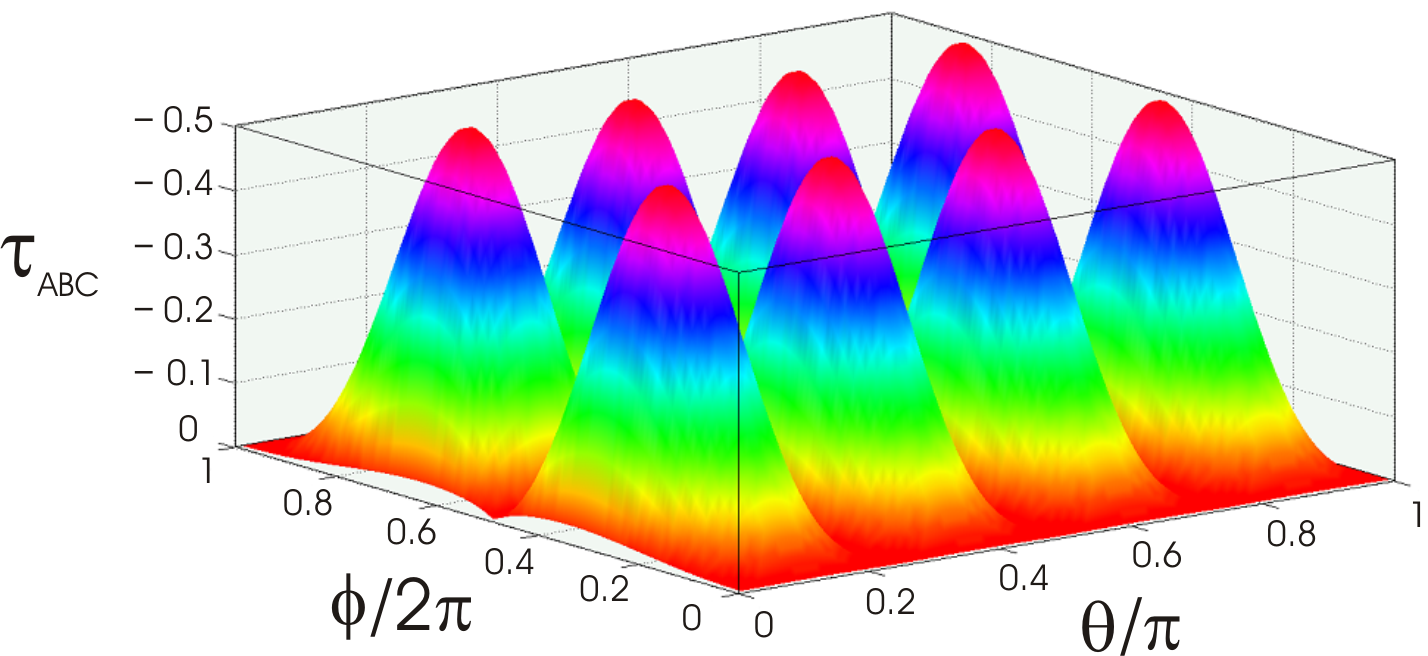} {}
\end{center}
\caption{(Color Online) We numerically evaluate the minus EOF tangle $-\tau_{ABC}$ as a function of $\theta$ and $\phi$, where both angles defined completely a $W$ state. One can notice that the EOF tangle is always lesser than zero for different $W$ states.}\label{tauW}
\end{figure}

We now extend the computation of the EOF tangle for a tripartite pure state with dimension  $2\times2\times N$. This system is particularly interesting since it allows to characterize how a two-qubit system becomes entangled with an environment. We can rewrite Eq. (\ref{tauabc}) as
\begin{equation*}
2\;\tau_{ABC}=\underbrace{\Delta^\leftarrow_{AB} + \Delta^\leftarrow_{AC}}_{2\tau_A} + \underbrace{\Delta^\leftarrow_{BA} + \Delta^\leftarrow_{BC}}_{2\tau_B} + \underbrace{\Delta^\leftarrow_{CA} + \Delta^\leftarrow_{CB}}_{2\tau_C}.\label{tauabc1}
\end{equation*}
The main obstacle to calculate $\tau_{ABC}$ is to determine the four elements $\Delta^\leftarrow_{AC}$, $\Delta^\leftarrow_{BC}$, $\Delta^\leftarrow_{CA}$, and $\Delta^\leftarrow_{CB}$, since the elements $\Delta^\leftarrow_{AB}$ and $\Delta^\leftarrow_{BA}$ can be computed straightforwardly. On the other hand, by examining Eq.~(\ref{conserv}), we notice that $\Delta^\leftarrow_{AC}=\Delta^\leftarrow_{BC}$ and both elements can be evaluated through the EOF and the mutual information of the pair $AB$, once $\Delta^\leftarrow_{AC}=\Delta^\leftarrow_{BC}=I_{AB}-2E_{AB}$.
Similarly, $\Delta^\leftarrow_{CA}=\Delta^\leftarrow_{BA}$, while $\Delta^\leftarrow_{CB}=\Delta^\leftarrow_{AB}$, and so the resulting EOF tangle is given by $\tau_{ABC}=\Delta^\leftarrow_{AB}+\Delta^\leftarrow_{BA}+\Delta^\leftarrow_{AC}$. 

\section{Sudden death of entanglement}
To exemplify we examine how two qubits evolve when coupled to two independent environments at zero temperature acting as amplitude damping channels. The effective system in question has a minimum dimension of $2\times 2 \times 4$ and, as exposed above, $\tau_{ABC}$ can be trivially calculated. The model Hamiltonian can be written as
\begin{equation*}
H=\omega^{(i)}_{0}\sigma^{(i)}_{+}\sigma^{(i)}_{-}+\sum_k\omega^{(i)}_{k}{a^{(i)}_{k}}^{\dagger}a^{(i)}_{k}+\left(\sigma^{(i)}_{+}B^{(i)}+\sigma^{(i)}_{-}{B^{(i)}}^{\dagger}\right),
\end{equation*}
 where $B^{(i)}=\sum_{k}g^{(i)}_{k}a^{(i)}_{k}$ with $g^{(i)}_{k}$ being the coupling
constant, $\omega^{(i)}_{0}$ is the transition frequency of the
i-$th$ qubit, and $\sigma^{(i)}_{\pm}$ are the i-$th$ qubit raising and
lowering operators. The index $k$ labels
the reservoir field modes with frequencies $\omega^{(i)}_{k}$, and
${a^{(i)}_{k}}^{\dagger}$ ($a^{(i)}_{k}$) is the usual creation
(annihilation) operator. Hereafter, the Einstein
convention sum is adopted. Moreover, we suppose that the environments are represented by a
bath of harmonic oscillators, and the spectral density is of the
form
$
J(\omega)=\gamma_{0}\lambda^{2}/2\pi[{(\omega_{0}-\omega)^{2}+\lambda^{2}}]$ where $\lambda$ is connected to the reservoir correlation time $\tau_{B}$
by the relation $\tau_{B}\approx1/\lambda$, and $\gamma_{0}$ is
related to the time scale $\tau_{R}$ over which the state of the
system changes, $\tau_{R}\approx1/\gamma_{0}$.  
\begin{figure}[htbp]
\begin{center}
\includegraphics[width=.5\textwidth]{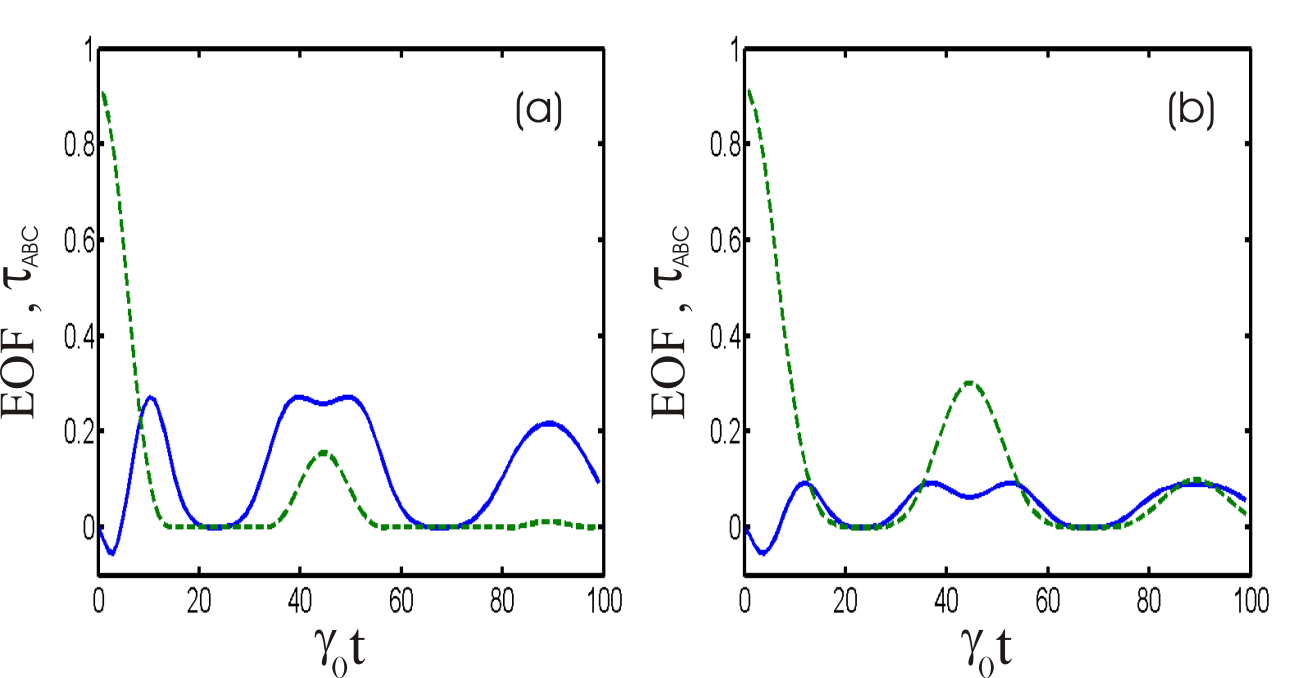} {}
\end{center}
\caption{(Color Online) The EOF dynamics between the qubits and the $\tau_{ABC}$ dynamics. The parameter adopted for the initial state in (a) $a=1/\sqrt{3}$, while in (b) we use $a=\sqrt{2/3}$. We suppose a strong coupling in both cases with $\lambda=0.01\gamma_0$.}\label{esd}
\end{figure}

{In Fig.~\ref{esd} (a) we consider an initial Bell-like state:}
$\left|\psi\right\rangle = a\left|00\right\rangle + \sqrt{1-a^2}\left|11\right\rangle$, with $a=1/\sqrt{3}$, which is a typical situation where the entanglement sudden death (SDE) phenomenon \cite{ESDc} occurs. On the other hand, in In Fig.~\ref{esd} (b) we consider the non-Markovian regime with $a=\sqrt{2/3}$ and one can see that the EOF goes to zero only for some discrete values of time. In contrast for a Markovian dynamics this last initial situation \textit{would not} show the ESD phenomenon.
 We see in both Figs.~\ref{esd} (a) and (b) that the tangle $\tau_{ABC}$ quickly becomes positive, meaning that the environment induces the global state to converge in one state that satisfies the monogamic relation. This characteristic occurs because the QD tends to decay while the classical correlation tends to be stable, at least after the transition where the classical correlation becomes greater than the quantum one. More importantly, we might establish a connection between the ESD phenomenon and the balance of classical and quantum correlations. To understand this characteristic, we must revisit Eq.~(\ref{mono1}).
When the entanglement between $AB$ vanishes, the entanglement between the qubit $A$ and the environment must be equal to $E_{A(BC)} + \tau_A$. Moreover, if the EOF \textit{does not} suddenly disappears, $E_{AB}=0$ just when $E_{AC}=E_{A(BC)}$ since the system $AC$ is pure and thus $\tau_A$ must vanish. In other words, in this situation the entanglement suddenly disappears when the {{\textit{discrepancy}}} in the global system (qubits plus environment) is zero, or when the information gain in using global operations is equal to the information acquired by using local operations in $B$ and $C$ only.

\section{Mixed tripartite states and the Squashed Entanglement}

To finalize we sketch an extension for a tripartite mixed state.
The EOF tangle can be positive or negative corresponding respectively to the GHZ or W entanglement.
In a mixed state, the ensemble may have elements of both families and our aim, in defining a tripartite measure, must be minimizing these two components over all the ensembles. For this purpose, we define two quantities 
\begin{eqnarray}
\tau_{GHZ}(|\psi\rangle)=\max\{0,\tau_{GHZ}\},\\
\tau_{W}(|\psi\rangle)=\max\{0,\tau_{W}\},
\end{eqnarray}
for any pure tripartite state $|\psi\rangle$. In this way, these measures accounts for the amount of GHZ and W entanglement in $|\psi\rangle$, Also, they are mutually exclusive in the sense that, if one is greater than zero, the other is zero. Therefore, for mixed, we can define $\tau$ as
\begin{equation}
\tau(\rho) = \min_{\cal E}\left\{\sum_i p_i (\tau_{GHZ}(|\phi_i\rangle)+\tau_W (|\phi_i\rangle))\right\},
\label{taumixed}
\end{equation}
where the minimization goes over all ensembles $\cal E$ such that $\rho=\sum_i p_i |\phi_i\rangle \langle \phi_i|$.

Now we discuss an issue related to mixed states with more general relevance for entanglement theory.
It is known that under some special conditions the EOF is equivalent to the Squashed Entanglement (SE) \cite{chrwinter}, a monogamous entanglement measure \cite{koashi2004}, but it is not clear that all the monogamous distribution of EOF is related to the SE as well. 
To investigate this relation we note that SE is defined
as \[E_{sq}(\rho_{AB}) = \frac{1}{2}\inf_{\varrho_{ABC}\in K}S(A:B | C),\]
where $K$ is the set of all density matrices $\rho_{ABC}:\; \rho_{AB}=Tr_C \rho_{ABC}$, and  $S(A:B|C)$ is the conditional mutual information for the extended system $\rho_{ABC}$.  It is known that the SE is upper bounded by the EOF, $E_{sq}(\rho_{AB})\le E_{AB}$. By Eq. (\ref{mono1}) for a pure $ABC$ system if EOF is monogamous,
\[E_{AB}+E_{AC}=E_{AB}+\delta^\leftarrow_{AB} + S(A|B)\le S_A,\]
and since $S_A=S(A:B)+S(A|B)$, we have
\[E_{AB}+\delta^\leftarrow_{AB} \le S(A:B),\] for $\rho_{ABC}$ pure, when $S(A:B|C)=S(A:B)$. In fact for arbitrary mixed states $S(A:B|C)$ can be larger or smaller than $S(A:B)$, depending on the kind of correlations allowed for the state. So it turns out that for arbitrarily mixed $\rho_{ABC}$, there are instances for which $S(A:B|C)\ge S(A:B)$ that certainly $S(A:B|C)\ge E_{AB}+\delta^\leftarrow_{AB}$. There are also other instances, for which $S(A:B|C)\le S(A:B)$, where either $S(A:B|C)\ge E_{AB}+\delta^\leftarrow_{AB}$ or $S(A:B|C)\le  E_{AB}+\delta^\leftarrow_{AB}$, such as the EOF can be larger or smaller than QD, in distinct instances. So when EOF is monogamic (and consequently the QD) there are instances, for arbitrarily mixed states $\rho_{ABC}$, where
\begin{equation}E_{sq}(\rho_{AB})\le\frac{1}{2}\left({E_{AB}+\delta^\leftarrow_{AB}}\right),\label{sq1}\end{equation} and other distinct instances where
 \begin{equation}E_{sq}(\rho_{AB})\ge\frac{1}{2}\left({E_{AB}+\delta^\leftarrow_{AB}}\right),\label{sq2}\end{equation} where we profited from the fact that the minimization over $C$ does not affect the right-hand side of the inequalities. To simplify the discussion we can check what occurs to states saturating both (\ref{sq1}) and (\ref{sq2}). There the SE is always related not only to the EOF, but to  the QD of the pair $AB$ as well. Interestingly when the reduced state $\rho_{AB}$ is pure we recover the well known results that the SE is equal to the EOF. But that is also true when there is full permutation invariance on $\rho_{ABC}$, $\delta^\leftarrow_{AB}=E_{AB}$, a fact not known previously. What is known is that when the set of states that minimize SE is constrained by states of the form $\rho_{ABC}=\sum_k|\psi_{AB}^k\rangle\langle\psi_{AB}^k|\otimes|k\rangle\langle k|$, it correspond to the ensemble minimization employed for the derivation of the EOF, and so the SE is equal to EOF. We see that this set corresponds to the states saturating both (\ref{sq1}) and (\ref{sq2}), only when there are total permutation invariance of $ABC$. 
 
 On a complementary side, when the EOF is not monogamic, $E_{AB}+\delta^\leftarrow_{AB} + S(A|B)> S_A$, and we always get that for arbitrarily mixed states $\rho_{ABC}$
 \[E_{sq}(\rho_{AB})<\frac{1}{2}\left({E_{AB}+\delta^\leftarrow_{AB}}\right),\] never attaining equality.

\section{discussion}
To conclude we have discussed an important and simple question: what is the amount of entanglement between $A$ and $BC$ that cannot be accounted for by the entanglement of $A$ with $B$ and $C$ separately? When the entanglement is measured by concurrence, CKW showed that this amount is always positive. On the other hand
for the EOF can assume either positive or negative values, depending on the gain that global operations give over local operations.
The reason for the negative signal of the EOF tangle, rather than being a failure of the accounting for a ``residual entanglement'',  is clarified in terms of the discrepancy between classical and quantum correlations. We have demonstrated the necessary conditions for this EOF tangle to be positive by considering an arbitrary pure tripartite system. Furthermore,  the   {\textit{discrepancy}}  exhibits a conservative relation because one particular subsystem shares always the same amount of  {\textit{discrepancy}} with other complementary subsystems. This description allows the study of the EOF tangle for a three qubits system in a very simple manner. The EOF tangle is always negative for the $W$ states family, while it is always positive for the $GHZ$ states given by Eq. (\ref{estghz}). Moreover, we described a simple numerical procedure to calculate the EOF tangle for an arbitrary tripartite system with dimension $2\times2\times N$, which enables the possibility of studying the EOF tangle between two qubits and an environment with $N$ degrees of freedom. In this sense, by examining two qubits coupled to independent amplitude damping channels, we show that the ESD phenomenon is intimately connected to the {\textit{discrepancy}} of the global system. 
We believe that this relation may allow a deep understanding of the distribution of entanglement and correlation in multipartite systems.

This work is supported by FAPESP and CNPq through the
National Institute for Science and Technology of Quantum
Information (INCT-IQ). MCO also acknowledges support by iCORE.

\textit{Note added:} While finishing this paper we became aware of related works \cite{ar1}. There the authors relate the EOF monogamy with the QD monogamy.


\begin{thebibliography}{99}
\bibitem{footnote1} Here extended is to be considered as a system given by the original systems $AB$ plus the ``extension" $C$, so that $\rho_{AB}= Tr_C \rho_{ABC}$.
\end{thebibliography}
\end{document}